\documentclass[aps,prb]{revtex4}

\newcommand{\be}{\begin{equation}}
\newcommand{\ee}{\end{equation}}
\newcommand{\bc}{\begin{center}}
\newcommand{\ec}{\end{center}}
\newcommand{\bea}{\begin{eqnarray}}
\newcommand{\eea}{\end{eqnarray}}
\newcommand{\ba}{\begin{array}}
\newcommand{\ea}{\end{array}}

\begin{document}

\title{Electroreflectance spectroscopy in self-assembled quantum dots: lens symmetry}

\author{A. H. Rodr\'{\i}guez$^{1,2}$, C. Trallero-Giner$^{1}$, Mart\'{\i}n Mu\~{n}oz$%
^{3,4}$, and Mar\'{\i}a C. Tamargo$^{3}$}

\affiliation{$^{1}$Facultad de F\'{i}sica, Universidad de La Habana, 10400 C. Habana,\\
Cuba.\\
$^{2}$Instituto de F\'{i}sica, Universidad Aut\'{o}noma de Puebla, Apdo.\\
Postal J-48, Puebla, Pue. 72570, M\'{e}xico.\\
$^{3}$Chemistry Department, City College of the City University of New York,%
\\
Convent Avenue at 138th Street, New York, New York 10031.\\
$^{4}$Physics Department, Virginia Commonwealth University, 1020 W. Main\\
Street Richmond, VA 23284.}

\begin{abstract}
Modulated electroreflectance spectroscopy $\Delta R/R$ \ of semiconductor
self-assembled quantum dots is investigated. The structure is modeled as
dots with lens shape geometry and circular cross section. A microscopic
description of the electroreflectance spectrum and optical response in terms
of an external electric field (${\bf F}$) and lens geometry have been
considered. The field and lens symmetry dependence of all experimental
parameters involved in the $\Delta R/R$ spectrum have been considered. Using
the effective mass formalism the energies and the electronic states as a
function of ${\bf F}$ and dot parameters are calculated. Also, in the
framework of the strongly confined regime general expressions for the
excitonic binding energies are reported. Optical selection rules \ are
derived in the cases of the light wave vector perpendicular and parallel to $%
{\bf F}$. Detailed calculation of the Seraphin coefficients and
electroreflectance spectrum are performed for the InAs and CdSe
nanostructures. Calculations show good agreement with measurements recently
performed on CdSe/ZnSe when statistical distribution on size is considered,
explaining the main observed characteristic in the electroreflectance
spectra.
\end{abstract}

\maketitle

\section{Introduction}

Modulation spectroscopy is a technique based on the changes of the
reflectivity of a sample when a periodic perturbation is applied. Due to its
nature, this technique provides derivative-like spectra related to the
optical transitions in the structure under consideration. Since its early
stages in the 60s\cite{cardona,cardona1,seraphin,aspnes,pollak} this
technique has been providing valuable information about the properties of
bulk/thin film semiconductors, reduced dimensional systems such as quantum
wells and superlattices,\cite{pollak1,glermbocki} and semiconductor device
structures.\cite{pollak2,pollak3} In spite of the versatility and success of
modulation spectroscopy few works have been done using this technique for
the analysis of quantum wires\cite{geddo} and quantum dot (QD) structures.
\cite{martin,qiang} The optical properties of self-assembled QDs (SAQD)\
have been widely studied using photoluminescence,\cite{medeiros,heitz}
photoluminescence excitation spectroscopy,\cite{medeiros,heitz} and time
resolved photoluminescence.\cite{raymond} However, the information obtained
is restricted to lower energy states and does not allow to study the shape
of the QD potential or the coupling effects in stacked structures. Even
though modulation spectroscopy allows to perform studies of lower and higher
energy transitions in QD structures very little work has been performed on
this subject.\cite{martin,qiang}

Reference [\onlinecite{martin}] reported contactless electroreflectance
(CER), which is a modulated technique that measures the changes in the
optical reflectance of the material with respect to a modulating electric
field, at room temperature in CdSe QDs with ZnSe barriers. The studied
structure consists of a GaAs buffer layer followed by GaAs/AlAs short-period
superlattice and CdSe QDs with ZnSe barriers (see inset in Fig. 1). The
corresponding spectrum shows a profile related to the buffer layer,
superlattices, QD region, wetting layer, and barriers. In the spectral
region $\hbar \omega <2.2$ eV the typical Franz Keldysh oscillations are
present, which are fit using Lorentzian broadened electro-optical functions.
\cite{pollak1} Also, the structures originating above 2.6 eV were fit using
the first derivative of a Gaussian lineshape.\cite{pollak4} As shown in Fig.
1 the electroreflectance spectrum coming from the confined QDs region
presents a broad structure which cannot be fit using standard
electro-optical functions.\cite{cardona} It is clear that a reliable
theoretical model for any modulation spectroscopy technique applied to
quantum dots should include the primordial geometric factor. The modulation
spectroscopy reflects doubtless the inherent quantum dot geometries.

In this paper we present a modulation spectroscopy study of quantum lens
(QL) structures, based on contactless electroreflectance (CER). By full
incorporation of the elements entering these experiments, namely the effects
of the geometry and electric field on excitonic states confined in the dot,
the oscillator strengths, exciton energies, and Seraphin coefficients
involved in the dielectric function, we provide the basis for quantitative
analysis of CER in SAQD with lens symmetry. We study the effects of lens
parameters on the {\em electroreflectance spectrum,} $\Delta R/R,$ and the
optical response. We found that the lens symmetry has strong and clear
signatures in the modulation spectroscopy, photoluminescence, and
photoluminescence excitation spectroscopy. Moreover, we show that a detailed
analysis of the optical response could provide information on the lens
geometry and effective mass of the carriers, since they affect strongly the
general features and overall peak distribution and amplitudes of the $\Delta
R/R$ profiles. Since the modulation spectroscopy data had played a prevalent
role in the study of III-V and II-VI semiconductors compounds, and because
of its intrinsic interest, we present a detailed analysis of a model for a
SAQD, which captures the essential physics of the problem. In fact, the lens
symmetry is likely to be a good model of SAQDs, where the characteristic
dimensions, height/diameter$<1$ are typical in these systems and our model
should provide a good description.

The remainder of the paper is organized as follows. Section II deals with
the general trends of the theory for $\Delta R/R$ applied to the case of a
quantum lens, discussing the nature of the excitonic states in these
structures and taking into consideration the external electric field
effects. Section III presents theoretical calculations for the InAs/GaAs
system, as well as the CdSe/ZnSe system. Section IV contains a fit to the $%
\Delta R/R$ data in CdSe/ZnSe SAQD structures. Finally, Sec. V\ is devoted
to the conclusions. In the appendix, we present the behavior of the Seraphin
coefficients for InAs/GaAs and CdSe/ZnSe quantum lenses and some technical
details of the calculations.

\section{Basic relations}

A precise knowledged of the interband transition energies in a semiconductor
can be traced by measuring the electroreflectance spectra.\cite{cardona}
This spectroscopy technique is based on the modulation of an ac external
electric field which modifies the shape of the dielectric function $%
\varepsilon (\hbar \omega ).$\cite{aspnes} For normal incidence of the light
the modulated electroreflectance, $\Delta R/R$ is related to the real, $%
\varepsilon _{1}$, and imaginary, $\varepsilon _{2},$ parts of the
dielectric function by

\begin{equation}
\frac{\Delta R(\hbar \omega ,F)}{R(\hbar \omega ,0)}=\alpha (\hbar \omega
)\,\Delta \varepsilon _{1}(\hbar \omega ,F)+\beta (\hbar \omega )\,\Delta
\varepsilon _{2}(\hbar \omega ,F),  \label{1}
\end{equation}
where $\Delta \varepsilon _{i}=\varepsilon _{i}(\hbar \omega ,F)-\varepsilon
_{i}(\hbar \omega ,0)$ ($i=1,2$), $F$ is the intensity of the electric
field, $\hbar \omega $ is the photon energy, and $\alpha ,$ $\beta $ are the
Seraphin coefficients (see Appendix A).

Using the standard semi-classical approach to describe the interaction
between light and matter, the imaginary part of the dielectric function for
direct allowed transitions takes the form

\begin{equation}
\varepsilon _{2}=16\pi \frac{a_{B}^{3}}{V_{o}}\frac{R_{y}^{2}}{(\hbar \omega
)^{2}m_{0}}\sum_{\alpha _{e},\alpha _{h}}\left| \int \Psi _{\alpha
_{e},\alpha _{h}}(r,r)d^{3}r\right| ^{2}\left| \widehat{{\bf e}}.{\bf p}%
_{cv}\right| ^{2}\frac{\gamma _{\alpha _{e},\alpha _{h}}}{(\hbar \omega
-E_{\alpha _{e},\alpha _{h}})^{2}+\gamma _{\alpha _{e},\alpha _{h}}^{2}},
\label{2}
\end{equation}
where $a_{B}$ is the Bohr radius, $R_{y}$ is the Rydberg constant, $V_{o}$
is the effective volume taking place in the process, $m_{0}$ is the free
electron mass, $\widehat{{\bf e}}$ is the polarization vector of the
incident light, ${\bf p}_{cv}$ is the interband optical matrix element
between conduction, $c,$ and valence, $v,$ bands, $\gamma _{\alpha
_{e},\alpha _{h}}$ is the broadening parameter of the Lorentzian function.
In the above equation $\left| \int \Psi _{\alpha _{e},\alpha
_{h}}(r,r)d^{3}r\right| ^{2}$ are the oscillator strengths for the allowed
interband transitions to the states $\Psi _{\alpha _{e},\alpha _{h}}$ with
energies $E_{\alpha _{e},\alpha _{h}}$.

The Kramers-Kronig relations provide the real part of dielectric function $%
\varepsilon _{1,}$ i.e.:

\begin{equation}
\varepsilon _{1}=1+16\pi \frac{a_{B}^{3}}{V_{o}}\frac{R_{y}^{2}}{(\hbar
\omega )^{2}m_{0}}\sum_{\alpha _{e},\alpha _{h}}\left| \int \Psi _{\alpha
_{e},\alpha _{h}}(r,r)d^{3}r\right| ^{2}\left| \widehat{{\bf e}}.{\bf p}%
_{cv}\right| ^{2}{\cal L(}\hbar \omega ,E_{\alpha _{e},\alpha _{h}}{\cal )},
\label{5}
\end{equation}
where

\begin{equation}
{\cal L}(\hbar \omega ,E_{\alpha _{e},\alpha _{h}})=\frac{E_{\alpha
_{e},\alpha _{h}}-\hbar \omega }{(\hbar \omega -E_{\alpha _{e},\alpha
_{h}})^{2}+\gamma _{\alpha _{e},\alpha _{h}}^{2}}\;+\frac{E_{\alpha
_{e},\alpha _{h}}+\hbar \omega }{(\hbar \omega +E_{\alpha _{e},\alpha
_{h}})^{2}+\gamma _{\alpha _{e},\alpha _{h}}^{2}}-\frac{2E_{\alpha
_{e},\alpha _{h}}}{E_{\alpha _{e},\alpha _{h}}^{2}+\gamma _{\alpha
_{e},\alpha _{h}}^{2}}.  \label{7}
\end{equation}
Two independent optical configurations are possible by choosing properly the
direction of the light wave vector ${\bf \kappa }$ with respect to the
applied electric field: i) ${\bf \kappa }$ $\Vert $ ${\bf F}$ $\Vert $ $%
\widehat{{\bf z}}$ and the vector of polarization $\widehat{{\bf e}}$ $\perp
$ $\widehat{{\bf z}}$ which is the typical configuration used in CER
experiments. Here, the three valence bands, lh, so and heavy-hole (hh) can
couple to the incident light. ii) ${\bf \kappa }$ perpendicular to ${\bf F}$
$\Vert $ $\widehat{{\bf z}}$ chosen along the quantum lens growth direction
and the vector polarization, $\widehat{{\bf e}}$ $\Vert $ $\widehat{{\bf z}}%
. $ In this case the light-hole (lh) and split-off (so) valence bands will
contribute to the optical spectrum.

\subsection{Electronic structure}

We will consider a typical SAQD with lens symmetry that presents a circular
cross section of radius $a$ and height $b.$ Electron-hole pairs (EHP) are
confined in the SAQD domain under an electric field $F$ parallel to its z
axial symmetry axis. The exciton wave functions are taken as solutions of
\begin{equation}
\left[ -\frac{\hbar ^{2}}{2m_{e}^{\ast }}\bigtriangledown _{e}^{2}-\frac{%
\hbar ^{2}}{2m_{h}^{\ast }}\bigtriangledown _{h}^{2}-e{\bf F}\cdot ({\bf r}%
_{e}-{\bf r}_{h})-\frac{e^{2}}{\epsilon \left| {\bf r}_{e}-{\bf r}%
_{h}\right| }\right] \Psi _{\alpha _{e},\alpha _{h}}({\bf r}_{e},{\bf r}%
_{h})=(E-E_{g})\Psi _{\alpha _{e},\alpha _{h}}({\bf r}_{e},{\bf r}_{h}),
\label{Hami}
\end{equation}
where $E_{g}$ is the gap energy, $\epsilon $ is the dielectric constant, and
$m_{i}^{\ast }$ (i=e,h) is the quasiparticle effective mass. In the strong
spatial confinement (electron-hole Coulomb interaction can be considered as
a perturbation) and according to the axial symmetry of the quantum lens, the
electron-hole pair wave function $\Psi _{\alpha _{e},\alpha _{h}}$ is given
by a product of electronic wave functions $\Psi _{i}({\bf \rho }_{i})\exp
(im_{i}\phi _{i}).$ Here, $m_{i}$ is the z component of the orbital angular
momentum and functions $\Psi _{i}$ satisfy a bi-dimensional Schr\"{o}dinger
equation for quantum dots with lens-shape geometry in an electric field.
Closed solutions of one-particle wave functions $\Psi _{N,m}$ and energy
levels $E_{N,m}$ ($N$ enumerates, for a fixed value of $m$, the electronic
levels by increasing value of the energy) as a function of the applied
electric field and lens shape geometry have been published elsewhere. \cite
{arezky,arezky1} The excitonic correction, appearing in Eq. (\ref{Hami}) has
been considered in first order perturbation theory. It is possible to
identify two cases:

i) $m_{e}=m_{h}=0,$ where the EH states are not degenerate and the excitonic
correction is directly given by

\begin{equation}
\Delta E_{0}=-\frac{2e^{2}}{\epsilon }\sum_{l=0}^{\infty }\frac{%
I_{l}(m_{e}=0,m_{h}=0;m^{\prime }=0)}{2l+1}.  \label{E0}
\end{equation}

ii) If $m_{e}=m_{h}\neq 0$ fourth-fold degeneracy of the EHP levels has to
be considered and a 4$\times $4 matrix for the exciton eingenvalue is
obtained. By symmetry it follows that the total z-component of the EHP
angular momentum $M=m_{e}+m_{h}$ is preserved under the electron-hole
correlation. For example, if $m_{e},m_{h}=\pm 1$ the states with $M=\pm 2$
are degenerate with an energy equal to

\begin{equation}
\Delta E_{0}=-\frac{2e^{2}}{\epsilon }\sum_{l=0}^{\infty }\frac{%
I_{l}(m_{e}=\pm 1,m_{h}=\pm 1;m^{\prime }=0)}{2l+1}.  \label{E1}
\end{equation}
While the states with $M=0$ are decupled with energies

\begin{eqnarray}
\Delta E_{+} &=&\Delta E_{0}+\frac{2e^{2}}{\epsilon }\sum_{l=0}^{\infty }%
\frac{I_{l}(m_{e}=1,m_{h}=-1;\left| m^{\prime }\right| =2)}{2l+1},
\label{E2} \\
\Delta E_{-} &=&\Delta E_{0}-\frac{2e^{2}}{\epsilon }\sum_{l=0}^{\infty }%
\frac{I_{l}(m_{e}=1,m_{h}=-1;\left| m^{\prime }\right| =2)}{2l+1}.
\label{E3}
\end{eqnarray}
$I_{l}$ are dimensionless functions given in the Appendix B. The energetic
order $\Delta E_{-}<\Delta E_{0}<\Delta E_{+}$ is preserved for any value of
the applied electric field or lens geometry (see Fig. 2). Notice that the
same behavior and equations are obtained for any values of the quantum
numbers $m_{e},m_{h}\neq 0$.

In Fig. 2 the first calculated excitonic energies $%
Eex(N_{e},m_{e};N_{h},m_{h})-E_{g}$ for CdSe quantum lens as a function of
the dimensionless electric field $F/F_{0}$ are plotted ($F_{0}=E_{0}/(|e|a),$
$E_{0}=\hbar ^{2}/(2\,m_{e}^{\ast }\,a^{2})$). For the calculations we have
used the values given in Table I. Two types of quantum lens are considered
representing the weak (Fig. 2a), $b/a=0.91$) and the strong (Fig. 2b), $%
b/a=0.51$) lens confinement domains, respectively. Excitonic states with $%
m_{e}=m_{h}=0$, $1,$ and $2$ are shown by solid, dashed, and dotted lines,
respectively. In both calculations the excitonic correction represents a
very small contribution to the total energy and the effect of the $F$ on $%
\Delta E$ is practically negligible. The splitting of the EHP levels, due to
the electron-hole correlation, diminishes as the confinement increases. In
the case of weak spatial confinement the electric field effect upon
energetic levels is stronger as shown in Fig. 2a) in comparison to Fig. 2b).
The interplay between $F$ and the ratio $b/a$ determines the peculiarities
of the excitonic energy as a function of $F$ in particular on the excited
states (for details see Ref. \onlinecite{arezky1}). Also, due to the lens
geometry, the electronic energies present an asymmetric Stark shift with the
external applied field.

\section{Electroreflectance}

In the following, we analyze the electroreflectance spectrum $\Delta R/R$
for the case of InAs/GaAs and CdSe/ZnSe SAQDs. This measurement gives rise
to sharp, differential-like spectra in the region of the transitions. In the
figures the main excitonic transitions are denoted by numbers (1,2,..),
which correspond to a particular set of quantum numbers $%
(N_{e},m_{e};N_{h},m_{h})$. Due to the axial symmetry, the interband
selection rules correspond to excitonic branch with $\Delta m=m_{e}-m_{h}=0$%
. The allowed transitions are resolved in the $\Delta R/R$ spectrum as
different ``effective gaps '' and the peak positions are directly
proportional to the lens geometry. In the case of $m_{e}=m_{h}\neq 0$ the
EHP degeneracy is broken and additional structure appears in the
electromodulation spectrum. We have only considered the incoming frequency
in the range below the energy barrier, according to the material parameters
listed in Table I. A full analysis of the electroreflectance response in
each system provides complementary information to photoluminescence and
photoluminescence excitation spectra to characterize the nanostructures
involved and the quantum lens geometry.

\subsection{InAs/GaAs}

Figure 3 displays the electroreflectance spectra of InAs dots embedded in GaAs
barriers for the cases of two independent optical configurations: a)
${\bf \kappa } \\ {\bf F}$ and $\widehat{{\bf e}}$ $\perp $ $%
\widehat{{\bf z}}$ and b) ${\bf \kappa \perp F}$ and $\widehat{{\bf e}}$ $%
\Vert $ $\widehat{{\bf z}}$. For the calculation the value of $Eg=1.51$ eV
for GaAs has been used. Solid vertical arrows show the excitonic transitions
for a lens geometry with $a=$16.0 nm and $b=$14.56 nm (solid lines), while
dashed vertical arrows correspond to a QL with $a=$20.5 nm, $b=$10.46 nm.
For the case of Fig. 3a) we used the corresponding Seraphin coefficient $%
\alpha $ and $\beta $ displayed in Fig. 6a), where the spatial confinement
effect it can be noticed . Also for closely spaced peaks, the interference
between different resonant levels increases and the $\Delta R/R$ signal is
not simply the result of single contributions.

Due to the relative oscillator strength of the hh and lh valence bands, the
electroreflectance features appear as relatively large resonant peaks in the case of
${\bf \kappa} || {\bf F}$ (Fig. 3a)) in comparison to
the spectrum for the optical configuration ${\bf \kappa \perp F,}$ $\widehat{%
{\bf e}}$ $\parallel $ $\widehat{{\bf z}}$ (Fig. 3b)). Labels 1 and 2 for
all graphs correspond to the transitions between $N_{h}=1,m_{h}=0\rightarrow
$, $N_{e}=1,m_{e}=0$ and $N_{h}=1,m_{h}=1\rightarrow $, $N_{e}=1,m_{e}=1,$
respectively. Notice in particular that in Fig. 3a) the transitions
involving the hh exciton are substantially stronger and the light hole
oscillator strength is about 10 times smaller than that corresponding to the
heavy hole. In the configuration where ${\bf \kappa }$ $\perp $ ${\bf F}$ $%
\Vert $ $\widehat{{\bf z}}$ and $\widehat{{\bf e}}$ $\parallel $ $\widehat{%
{\bf z}}$ (Fig. 3b)) the heavy hole excitonic branches are forbidden and
labels 1 and 2 represent the light hole contributions.

\subsection{CdSe/ZnSe}

To illustrate the role of the II-VI materials that compose a QL, Fig.\ 4
shows the electroreflectance spectrum as a function of the photon energy for
CdSe dots with ZnSe barriers. A value of $Eg=2.7$ eV for ZnSe is used for
the numerical evaluation. The obtained spectra correspond to the cases of
Fig. 2, solid line for a QL with $a=15.0$ nm $b=13.65$ nm, while dashed
lines to the geometry with $a=20.0$ nm, $b=10.20$ nm. To calculate $\Delta
R/R$ we used the $\alpha $ and $\beta $ parameters shown in Fig. 6b). The
stronger oscillation strength is due to the excitonic branch $N_{e}=N_{h}$
and $m_{e}=m_{h}$, the rest of the allowed transitions are too weak to be
resolved in ER spectrum. According to this, the exciton dispersion relations
calculated in Fig. 2 closely\ follow the calculated $\Delta R/R$ structure.
In general the spectra show the same general trend with respect to the InAs
case. Nevertheless, two main differences are present: i) Due to the heavy
hole mass and for a given geometry the electroreflectance spectrum of CdSe
has more structure that in the InAs QL. ii) The exciton degeneracy is broken
for $m_{e}=m_{h}\neq 0$ (displayed in Fig 4 as a circle for the case of weak
confined $b/a=0.91$). Excitonic binding energies $\Delta E,$ for II-VI
semiconductors are larger than the III-V ones and consequently, the exciton
degeneracy can be easily resolved by a spectroscopy technique.

\section{Application to Cadmium Selenide/Zinc Selenide quantum lens}

Figure 1 shows the experimental CER spectrum of CdSe/ZnSe SAQD at room
temperature.\cite{martin} We have performed calculations of the $\Delta R/R$
within the framework of the model developed in this paper, in order to
compare its ability to reproduce the experimental data. A typical QD
structure is shown in the inset of Fig. 1. Details about the growth
conditions of the CdSe/ZnSe QD samples are provided elsewhere.\cite{martin}
The CER spectra of these structures have been obtained using a
condenser-like system\cite{pollak4} consisting of a front wire grid
electrode with a second metal electrode separated from the first electrode
by insulating spacers, which are approximately 0.1 mm larger than the sample
dimension. The sample was placed between these two capacitor plates and the
electromodulation was achieved by applying an ac voltage of 1.2kV, 200 Hz
across the electrodes. In Fig. 1 we can identified the differential-like
spectra originating from the QDs in the region $2.2<\hbar \omega <2.5$ eV.
In Fig. 5 electroreflectance data for the CdSe QD are displayed as solid
circles. To fit the transitions originating from the QDs with QL geometry we
took $a_{0}=11.98$ nm and $b_{0}/a_{0}=0.24$. The solid line corresponds to
the evaluation of Eq. (\ref{1}) for a single quantum lens in presence of an
electric field equal to $F=50$ kV/cm and a constant exciton broadening
parameter $\gamma _{\alpha _{e},\alpha _{h}}=8$ meV. The sharp
differential-like structure matches very well with the measured QL allowed
optical transitions. These peaks correspond to electron-heavy hole
transitions $N_{e}=1,m_{e}=0\rightarrow N_{hh}=1,m_{hh}=0$ and $%
N_{e}=1,m_{e}=1\rightarrow N_{hh}=1,m_{hh}=1.$

Self-assembled\ quantum dots have a distribution on size and shape. For a
given photon energy $\hbar \omega $ we have to take into account the
contribution of all quantum lenses that fulfil the resonance conditions $%
\hbar \omega =E_{\alpha _{e},\alpha _{h}}$ and evaluate the average CER. In
our calculation we fixed the ratio $a_{0}/b_{0}$ and performed an average
over the size $a.$ The corresponding expression for the average
electroreflectance $\overline{\left( \Delta R/R\right) }$ is written as
\begin{equation}
\overline{\left( \frac{\Delta R}{R}\right) }=\int F(a)\frac{\Delta R}{R}da,
\label{eq:5}
\end{equation}
where a Gaussian size distribution function $F(a)$ with mean value $a_{0}$,
and FWHM $\sigma $ is assumed. Figure 5 displays our theoretical
calculations for the average $\overline{\left( \Delta R/R\right) }$ spectrum
(dashed lines) of an ensemble of CdSe quantum lens with average ratio $%
b_{0}/a_{0}=0.24,$ $a_{0}$=11.98 nm, and $\sigma =0.4$ nm$.$ It can be seen
that the observed measured broad spectrum is explained by size distribution
of the QLs. Hence, the $\Delta R/R$ signal of Fig. 5 is the contributions of
QLs in different resonance regimes, i.e. those excitonic transitions
fulfilling the condition $\hbar \omega =E_{n_{e},m_{e};n_{h},m_{h}}(a,b).$
Resonances with higher exciton states occur for larger $a$ and $b$ values
but are quenched by the size distribution function $F(a)$ present in Eq. (%
\ref{eq:5}).

\section{Conclusions}

The present theoretical description can be used to evaluate the modulated
electroreflectance spectra of III-V and II-VI SAQD with lens shape geometry.
Optical responses and electroreflectance spectra as a function of the
electric filed have been calculated for incoming photon energy above the
fundamental effective gaps in QLs semiconductors. The Seraphin coefficients
present a series of thresholds according to the excitonic ($N_{e},m_{e}$;$%
N_{h},m_{h})$ branch and the allowed optical transitions in the lens. The ER
for InAs and CdSe dots show sharp differential-like spectra which identify
the interband excitonic transitions of the QDs. The calculated $\Delta R/R$
dependence on $\hbar \omega $ reproduce quite well the experimental data
available for CdSe/ZnSe quantum dots. This fact indicates that the present
theoretical model through out this paper contains the main ingredients of
the electroreflectance spectroscopy in SAQD with lens geometry, and thus can
be used, in combination with experimental data, to obtain information on the
lens shape and other physical parameters related to the growth conditions of
the sample. An important outcome of the work is that by fitting experimental
data to this model we can estimate the size distribution of the QDs in the
capped structures, which is a parameter not easily determined by other means.

\appendix

\section{Seraphin coefficients}

These coefficients are related to the dielectric constant at zero electric
field. Their spectral dependences are obtained by the expressions
\begin{equation}
\alpha (\hbar \omega )=\frac{2n}{n^{2}+k^{2}}\frac{n^{2}-3k^{2}-1}{\left[
(n+1)^{2}+k^{2}\right] \left[ (n-1)^{2}+k^{2}\right] },  \label{A1}
\end{equation}
\begin{equation}
\beta (\hbar \omega )=\frac{2k}{n^{2}+k^{2}}\frac{3n^{2}-k^{2}-1}{\left[
(n+1)^{2}+k^{2}\right] \left[ (n-1)^{2}+k^{2}\right] }.  \label{A2}
\end{equation}
The refractive index, $n,$ and the extinction coefficient, $k,$ are
functions of $\varepsilon _{1}$ and $\varepsilon _{2}$ according to
\begin{equation}
n=\sqrt{\frac{\sqrt{\varepsilon _{1}^{2}+\varepsilon _{2}^{2}}+\varepsilon
_{1}}{2}}+n_{\infty },\;k=\sqrt{\frac{\sqrt{\varepsilon _{1}^{2}+\varepsilon
_{2}^{2}}-\varepsilon _{1}}{2}},  \label{n}
\end{equation}
where $n_{\infty }$ is the refractive index at high frequency. Inserting
Eqs. (\ref{2})- (\ref{4}) at $F=0$ into Eqs. (\ref{A1}), (\ref{A2}), and (%
\ref{n}) we obtain the values of the Seraphin coefficients for the optical
geometry ${\bf \kappa }$ $\Vert $ ${\bf F}$ $\Vert $ $\widehat{{\bf z}}$ and
$\widehat{{\bf e}}$ $\Vert $ $\widehat{{\bf x}}$. Coefficients $\alpha $ and
$\beta $ as a function of the photon energy for the InAs/GaAs and CdSe/ZnSe
QLs are shown in Figs. 6a) and Fig.6b), respectively. We have considered the
same lens geometries indicated in Figs. 3 and 4. Weak and strong spatial
confinement regimes are indicated by solid and dashed lines, respectively.
Labels 1 and 2 represent the optical transitions between states $N_{h}=1$, $%
m_{h}=0$ to $N_{e}=1$, $m_{e}=0$ and $N_{h}=1$, $m_{h}=1$ to $N_{e}=1$, $%
m_{e}=1$. From \ the figures the strong influence of the lens geometry on
the Seraphin coefficients and in consequence, on the electroreflectance, is
clear.

\section{Exciton matrix elements}

In Eqs. (\ref{E0})-(\ref{E3}) the matrix element $I_{l}$ has an explicit
expression (see Refs.\onlinecite{arezky1} and \onlinecite{jakson}):

\begin{eqnarray}
I_{l}(m_{e},m_{h};m^{\prime }) &=&\sum_{i,j,i^{\prime },j^{\prime }}^{\infty
}C_{i}(N_{e},m_{e})\,C_{j}(N_{h},m_{h})C_{i^{\prime
}}(N_{e},m_{e})\,C_{j^{\prime }}(N_{h},m_{h})  \nonumber \\
&&\,\left\langle f_{i,m_{e}}^{(o)}f_{j,m_{h}}^{(o)}\left| \frac{r_{<}^{l}}{%
r_{>}^{l+1}}P_{l}^{|m^{\prime }|}(\cos \theta _{e})P_{l}^{|m^{\prime
}|}(\cos \theta _{h})\right| f_{i^{\prime },m_{e}}^{(o)}f_{j^{\prime
},m_{h}}^{(o)}\right\rangle ,  \label{B1}
\end{eqnarray}
where coefficients $C_{i}(N,m)$ and functions $f_{i,m}^{(o)}$ are defined in
Ref. \onlinecite{arezky1}. $I_{l}$ depends on the lens deformation $b/a$ and
dimensionless electric field $F/F_{0}.$ The excitonic correction integrals
in Eq. (\ref{B1}) were obtained by a Monte Carlo algorithm over the
2-dimensional lens domain.\newline

{\large Acknowledgments. }This work was partially supported by the National
Science Foundation under Grant No. ECS0217646.

\begin{table}[tbp]
\caption{Parameters used in calculations.}
\begin{center}
\begin{tabular}{ccc} \hline\hline
Parameters & InAs & CdSe
\\ \hline
$E_g$ (eV) & 0.45$^a$ & 1.692$^b$ \\ \hline $\epsilon$ & 14.6$^a$ & 9.3$^a$ \\ \hline
$m_{e}^{\ast }/m_0$ & 0.023$^a$ & 0.11$^b$ \\ \hline $m_{hh}^{\ast }/m_0$ & 0.34$^a$
& 0.44$^b$ \\ \hline $m_{lh}^{\ast }/m_0$ & 0.027$^a$ & -- \\ \hline $\Delta E_c$
(\%) & {40\%}$^a$ & {85\%}$^b$ \\ \hline $\Delta E_v$ (\%) & {60\%}$^a$ & {15\%}$^b$
\\ \hline
$n_{\infty }$ & 3.517$^c$ & 2.5$^d$ \\ \hline $P^2/m_{0}$ (eV) & 10.0 $^c$ & 11.1
$^c$ \\ \hline $\gamma_{hh}$ (meV) & 3 & 3 \\ \hline$\gamma_{lh}$ (meV) & 5 & - \\
\hline\hline
\end{tabular}
\end{center}
\par
$^a$ Ref.\ [\onlinecite{eduardo}]\newline
$^b$ Ref.\ [\onlinecite{martin}]\newline
$^c$ Ref.\ [\onlinecite{Landolts}]\newline
$^d$ An average of the values reported in Ref.\ [\onlinecite{Landolts}]
\end{table}

\begin{figure}
\caption{Contactless electroreflectance spectrum $\Delta R(\hbar \protect%
\omega ,F)/R(\hbar \protect\omega ,0)$ for a CdSe quantum dot structure.
Solid line correspond to the spectral measurement. Dashed lines in the
spectral region $\hbar \protect\omega <2.2$ eV represent fit using standard
electro-optical functions, while for the region above 2.6 eV a fit using the
first derivative of a Gaussian lineshape. The QDs spectral region addressed
in this paper is $2.2<\hbar \protect\omega <2.5$. The inset shows a typical
CdSe/ZnSe sample studied.}
\label{fig1}
\end{figure}

\begin{figure}
\caption{Excitonic energy levels $Eex(N_{e},m_{e};N_{h},m_{h})-E_{g}$ for
CdSe quantum lenses as a function of the electric field. Excitonic branches
are labeled by $N_{e};N_{h}$ for the allowed optical transitions $\Delta
m=m_{e}-m_{h}=0$. a) Lens domain with $a=15$ nm, $b=$13.65 nm. b) $a=20$ nm,
$b=10.2$ nm.}
\label{fig2}
\end{figure}

\begin{figure}
\caption{Electroreflectance spectrum for InAs quantum lenses. Solid lines
correspond to the lens domain $a=16$ nm, $b=14.56$ nm and dashed lines to $%
a=20.5$ nm, $b=10.46$ nm. Optical configurations: a) ${\bf \protect\kappa }$
$\perp $ ${\bf F}$ $\Vert $ $\widehat{{\bf z}}$ and $\widehat{{\bf e}}$ $%
\Vert $ $\widehat{{\bf z}}$. b) ${\bf \protect\kappa }$ $\Vert $ ${\bf F}$ $%
\Vert $ $\widehat{{\bf z}}$ and $\widehat{{\bf e}}$ $\perp $ $\widehat{{\bf z%
}}$. In the calculation a value of $F=50$ kV/cm is used. Allowed excitonic
optical transitions are indicated by arrows}
\label{3}
\end{figure}

\begin{figure}
\caption{Electroreflectance spectrum for CdSe quantum lenses for the optical
configuration ${\bf \protect\kappa }$ $\Vert $ ${\bf F}$ $\Vert $ $\widehat{%
{\bf z}}$ and $\widehat{{\bf e}}$ $\perp $ $\widehat{{\bf z}}$. Solid lines
correspond to the lens domain $a=16$ nm, $b=14.56$ nm and dashed lines to $%
a=20$ nm, $b=10.2$ nm. $F=50$ kV/cm is used and the allowed excitonic
optical transitions are indicated by arrows.}
\label{4}
\end{figure}

\begin{figure}
\caption{Contactless electroreflectance spectrum $\Delta R(\hbar \protect%
\omega ,F)/R(\hbar \protect\omega ,0)$ for the CdSe/ZnSe quantum dot
structure shown in Fig. (1). Dots are the experimental data. The solid line
represents the calculation for a QL with radius $a_0=11.98$ nm and height $%
b_0= 2.88$ nm, while dashed lines correspond to average size calculation of $%
\overline{\Delta R(\hbar \protect\omega ,F)/R(\hbar \protect\omega ,0)}$ . }
\label{fig5}
\end{figure}

\begin{figure}
\caption{Seraphin coefficients $\protect\alpha $ and $\protect\beta $ for a
quantum lens in the optical configuration ${\bf \protect\kappa }$ $\Vert $ $%
{\bf F}$ $\Vert $ $\widehat{{\bf z}}$ and $\widehat{{\bf e}}$ $\perp $ $%
\widehat{{\bf z}}$. The allowed excitonic optical transitions are indicated
by arrows. a) InAs/GaAs. b) CdSe/ZnSe. }
\label{6}
\end{figure}


\begin{references}
\bibitem{volume}  The volume of a quantum lens is given by $V=\pi
a^{3}(b_{0}/a_{0}+(1/3)(b_{0}/a_{0})^{3})/2$ and for fixed ratio of $%
b_{0}/a_{0}$ the distribution in size $a$ is directly related to a
distribution in volume.
\end{references}
\end{document}